\documentclass{appolb}
\usepackage{graphicx}
\usepackage[utf8]{inputenc}
\usepackage{amsmath,amssymb,mathtools,commath}
\usepackage{bm,dsfont}
\usepackage{combelow}
\usepackage{ulem}
\usepackage{xcolor}

\newcommand{\epn}{\frac{d E_\perp^{(0)}}{d \eta}}

\newcommand{\xT}{\mathbf{x_\perp}}
\begin{document}
% \eqsec  % uncomment this line to get equations numbered by (sec.num)
\title{Development of transverse flow for small and large systems in conformal kinetic theory
\thanks{Presented at Quark Matter 2022 by C. Werthmann}%
% you can use '\\' to break lines
}
\author{Victor E. Ambru\cb{s}
%\address{Institut f\"ur Theoretische Physik, Johann Wolfgang Goethe-Universit\"at, Max-von-Laue-Strasse 1, D-60438 Frankfurt am Main, Germany}
\address{Department of Physics, West University of Timi\cb{s}oara, \\
Bd.~Vasile P\^arvan 4, Timi\cb{s}oara 300223, Romania}
%\address{Facultatea de Fizică, Universitatea de Vest din Timi\cb{s}oara, \\
%Bd.~Vasile P\^arvan 4, Timi\cb{s}oara 300223, Rom\^ania}
\\[3mm]
{S.~Schlichting, \underline{C.~Werthmann}
\address{Fakultät für Physik, Universität Bielefeld, D-33615 Bielefeld, Germany}
}
}
\maketitle
\begin{abstract}
We employ an effective kinetic description to study the space-time dynamics and development of transverse flow of small and large collision systems. By combining analytical insights in the few interactions limit with numerical simulations at higher opacity, we are able to describe the development of transverse flow from very small to very large opacities, realised in small and large collision systems. Surpisingly, we find that deviations between kinetic theory and hydrodynamics persist even in the limit of very large interaction rates, which can be attributed to the presence of the early pre-equilibrium phase. 
\end{abstract}
  
\section{Introduction}

Modern frameworks for simulations of heavy ion collisions typically feature an extended hydrodynamic phase that dominates the spacetime evolution of the system. This is where most of the commonly observed signatures of thermalization and colletive behaviour emerge, notably also transverse flow, which refers to anisotropies in the transverse momentum spectrum of measured particles, that build up in response to spatial anisotropies in the initial state. 

However, the picture is not as clear in the case of small systems. If the system is too dilute or if the gradients are too large, hydrodynamics may lose its applicability. Still, some signatures of collective behaviour, most notably transverse flow, are also experimentally observed in small systems, which calls for an alternative description. Another way to model the dynamics of a hadronic collision is via kinetic theory, a model describing the time evolution of phase space distributions based on microscopic interactions. This description is applicable on the whole range from very small to very large systems, with hydrodynamics being its limit for large interaction rates.

This motivates the work that we present here. To test the applicability of hydrodynamics, we employ a simplified kinetic theory description for the full opacity range and perform comparisons to hydrodynamics on the basis of transverse flow and related observables.

\section{Model and Setup}

We consider a simplified boost-invariant system of massless particles with an initially vanishing longitudinal pressure and transverse momentum anisotropy.  If we describe only energy-weighted degrees of freedom, the initial condition is then fully determined by the initial energy density $\epsilon(\tau_0,\xT)$. We choose the initial condition to be of a simple gaussian form, introducing only one anisotropic mode with $n-$fold rotational symmetry, whose magnitude is characterized by the eccentricity $\epsilon_n$ (see Refs.~\cite{Ambrus:2021fej,Kurkela:2020wwb} for details).

In our simplified kinetic theory description, the system evolves according to the relativistic Boltzmann equation in the relaxation-time approximation (RTA),
\begin{align}
    p^\mu \partial_\mu f = C_{\rm RTA}[f]=-\frac{p^\mu u_\mu}{\tau_R}(f-f_{\rm eq})\;,\label{eq:Boltzmann}
\end{align}
where $f$ is the distribution function, $p^\mu$ is the particle momentum, $f_{\rm eq} = (e^{p \cdot u/T} - 1)^{-1}$ is the Bose-Einstein equilibrium distribution and $u^\mu$ is the fluid four-velocity, %We choose the relaxation time $\tau_R$ to be that of a conformal system, 
%$\tau_R=5\eta/sT^{-1}$, where $T$ is the local temperature and $\eta/s$ is its specific shear viscosity.
$\tau_R=5\eta/sT^{-1}$ is the relaxation time, $T$ is the local temperature and $\eta/s$ the specific shear viscosity.

%It has been shown~\cite{Kurkela:2019kip} that in
In
conformal RTA, the system's time evolution depends only on one parameter, the opacity $\hat{\gamma}$, % which can be computed from the input parameters 
%as follows 
defined as
\cite{Kurkela:2019kip}
\begin{align}
    \hat{\gamma} =\left(5\frac{\eta}{s}\right)^{-1}\left(\frac{30}{\pi^2 \nu_{\mathrm{eff}}}\frac{1}{\pi R^2} \epn \right)^{1/4}R^{3/4}\;.
\end{align}
Here, $dE^{(0)}_\perp / d\eta$ is the initial energy per rapidity, $R$ is the system size and $\nu_{\rm eff}$ is the effective number of degrees of freedom. For $\eta/s=0.16$, the opacity takes typical values of $0.88$ in minimum-bias pp collisions and $9.2$ in central PbPb collisions. 
% Longitudinal cooling can be observed in terms of the decrease of transverse energy, computed as
% \begin{equation}
%  \frac{dE_\perp}{d\eta}=\int d^2x_\perp \int d^3p~p_\perp~f\;.
% \end{equation}
% The transverse flow properties are traditionally characterized in terms of the harmonic response coefficients $v_n$. In this work, we focus on the energy-weighted $v_n^E$ coefficients, defined as
% \begin{equation}
%  v_n^{E}=\frac{\int d^2x_\perp \int d^3p ~p_\perp~e^{in\phi_{p}}~f}{\int d^2x_\perp \int d^3p~p_\perp~f}\;,
% \end{equation}
% {\color{blue} For simplicity, we take the initial energy density $\epsilon_0(\mathbf{x}_\perp) = \frac{1}{\tau_0} dE_\perp^{(0)} / d\eta d^2x_\perp$ as a background Gaussian with a single harmonic eccentricity of magnitude $\epsilon_n$ (see Refs.~\cite{Kurkela:2020wwb,Ambrus:2021fej} for details).
% }

We will present numerical results for kinetic theory from a Relativistic Lattice Boltzmann approach~\cite{Ambrus:2018kug}. The hydrodynamic results that we will compare to are obtained using the vHLLE code~\cite{Karpenko:2013wva} with transport coefficients matched to RTA (see Ref.~\cite{Ambrus:2021fej} for details).

\section{Time evolution and opacity dependence in kinetic theory}

\begin{figure}[htb]
\centerline{%
\begin{tabular}{cc}
\includegraphics[width=.425\textwidth]{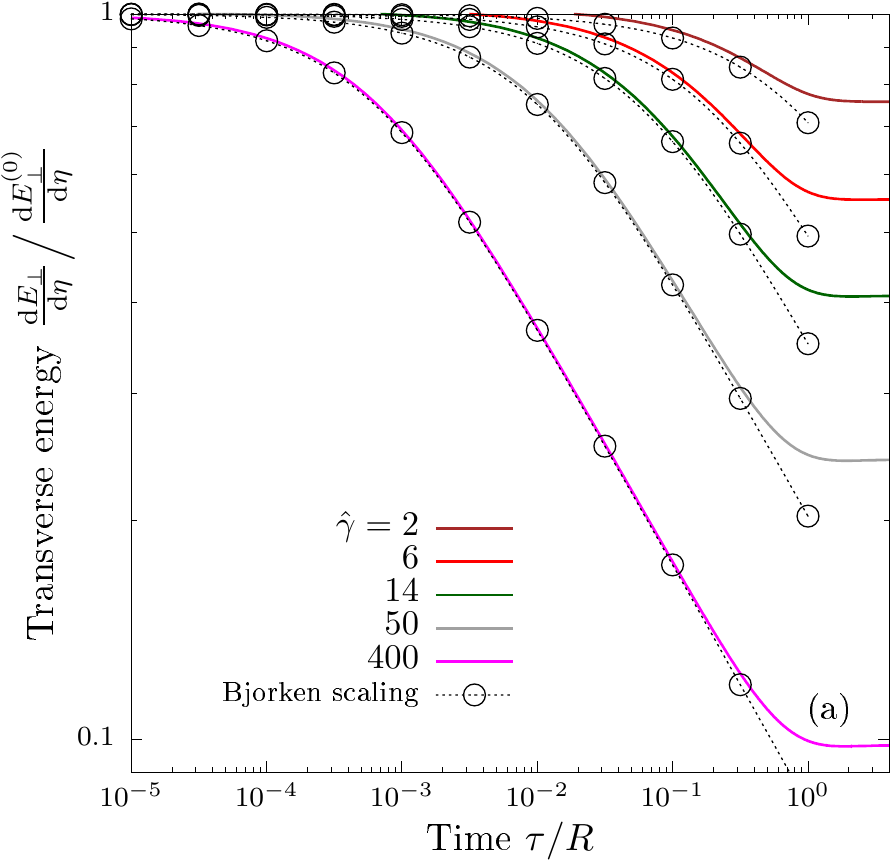} &
\includegraphics[width=.575\textwidth]{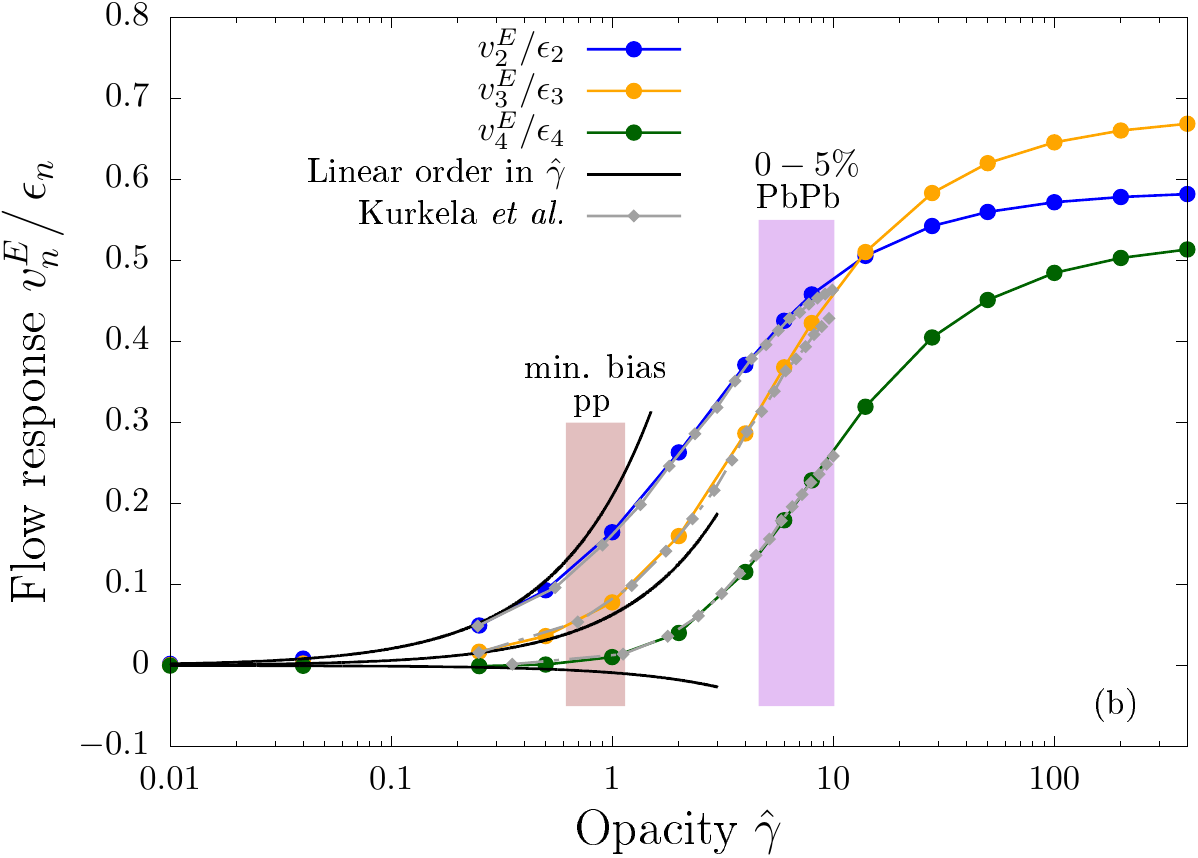} 
\end{tabular}
}
\caption{
(a) Time dependence of the transverse energy $dE_\perp / d\eta$ for various opacities. Our numerical results are shown with continuous lines and symbols, while the Bjorken scaling estimation is shown with dotted lines and empty circles.
(b) Opacity dependence of the flow responses $v_n^E/\epsilon_n$ obtained from numerical simulations (colored lines), using a linear order approximation (black lines), and from Ref.~\cite{Kurkela:2020wwb} (grey lines). 
%The large opacity limits indicated by the arrows were obtained by a constant plus power law fit. 
}
\label{fig:development}
\end{figure}

We characterize the transverse-plane dynamics using the transverse energy $dE_\perp / d\eta$ and the energy-weighted flow harmonics $v_n^E$, computed as
\begin{equation}
 \frac{dE_\perp}{d\eta}=\tau \int d^2x_\perp \int d^3p~p_\perp~f\;, \qquad
 v_n^{E}=\frac{\int d^2x_\perp \int d^3p ~p_\perp~e^{in\phi_{p}}~f}{\int d^2x_\perp \int d^3p~p_\perp~f}\;,
 \label{eq:obs_kin}
\end{equation}
where $(p_\perp,\phi_p)$ characterize the transverse-plane projection $\mathbf{p}_\perp$ of $p^\mu$. 

We examined these observables on the full parametric range in opacity and eccentricity~\cite{Ambrus:2021fej}. Here we present a selection of interesting results. Figure~\ref{fig:development}(a) shows the time evolution of $dE_\perp / d\eta$ for various values of $\hat{\gamma}$. At small values of $\tau / R$, the system is in the free-streaming regime and $dE_\perp / d\eta$ remains constant. During equilibration, longitudinal pressure is generated and $dE_\perp / d\eta$ is decreased due to work done by the system against longitudinal expansion. Since the transverse expansion is neglibile at early times, the longitudinal cooling of the system can be effectively described by 0+1D Bjorken dynamics locally at each point in the transverse plane. Corresponding results shown with dashed gray lines and empty circles, are in excellent agreement with the full numerical simulations. Finally, when $\tau / R \gtrsim 1$, transverse expansion becomes dominant and the cooling stops. 

In terms of the harmonic response coefficients, we present results for the opacity dependence of the response coefficients $v_n^E / \epsilon_n$ to a small initial-state harmonic anisotropy of eccentricity $\epsilon_n$. Our numerical results shown in Fig.~\ref{fig:development}(b) indicate that for small values of $\hat{\gamma}$, these coefficients scale linearly with the opacity $\hat{\gamma}$. Linearized results of $v_2^E$ and $v_3^E$ are good approximations for pp opacities. The numerical coefficients are reported in Eqs.~(86)--(88) of Ref.~\cite{Ambrus:2021fej}. In the intermediate opacity range, we validated our results by comparison with the results reported in Ref.~\cite{Kurkela:2020wwb} in an identical setup. Finally, at opacities several times larger than those of central PbPb collisions, the late-time values of $v_n^E / \epsilon_n$ saturate. We will examine this further in the next section.

% We also examined a large variety of linear and non-linear flow responses on the whole parameter range in opacity and eccentricity, the results of which are presented in~\cite{Ambrus:2021fej}. Fig.~\ref{fig:development}(b) shows our results for the late time limit of the linear flow responses $v_n/\epsilon_n$.
% We also show analytical results to linear order in opacity, which are tangential to the numerical curve and provide a good approximation at small opacities. 
% We were able to extend the analysis to very small and very large opacities, which is the regime where one could hope to make comparisons to hydrodynamics. And indeed, we do see that the response curves saturate at opacities $\hat{\gamma}\gtrsim 100$, indicating that the system reaches the ideal fluid limit.

\section{Comparing results from kinetic theory and hydrodynamics}

For a better understanding of the large $\hat{\gamma}$ limit of our kinetic theory results, we performed a comparison with hydrodynamic simulations. In order to facilitate this comparison, in this section we will consider observables that can be computed directly from the energy-momentum tensor $T^{\mu\nu}$,
\begin{equation}
 \frac{dE_{\rm tr}}{d\eta} = \tau \int d^2x_\perp (T^{xx} + T^{yy}), \quad 
 \epsilon_p = \frac{\int d^2x_\perp\, (T^{xx} - T^{yy} + 2iT^{xy})}{\int d^2x_\perp (T^{xx} + T^{yy})},\label{eq:obs_hydro}
\end{equation}
where $dE_{\rm tr} / d\eta$ and $\epsilon_p$  act as proxies for $dE_\perp / d\eta$ and $v_2$.

\begin{figure}
\centerline{
\begin{tabular}{cc}
\includegraphics[width=.5\linewidth]{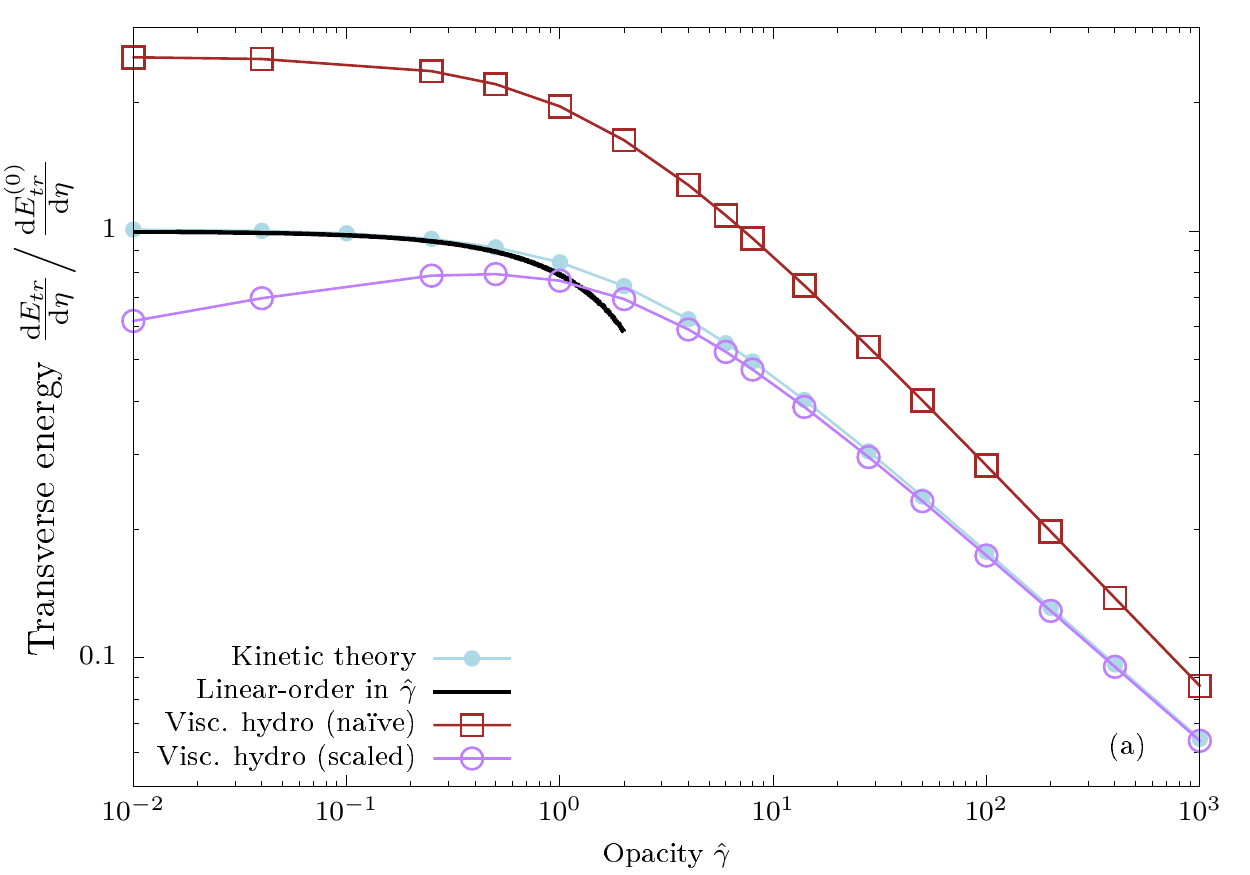} & 
\includegraphics[width=.5\linewidth]{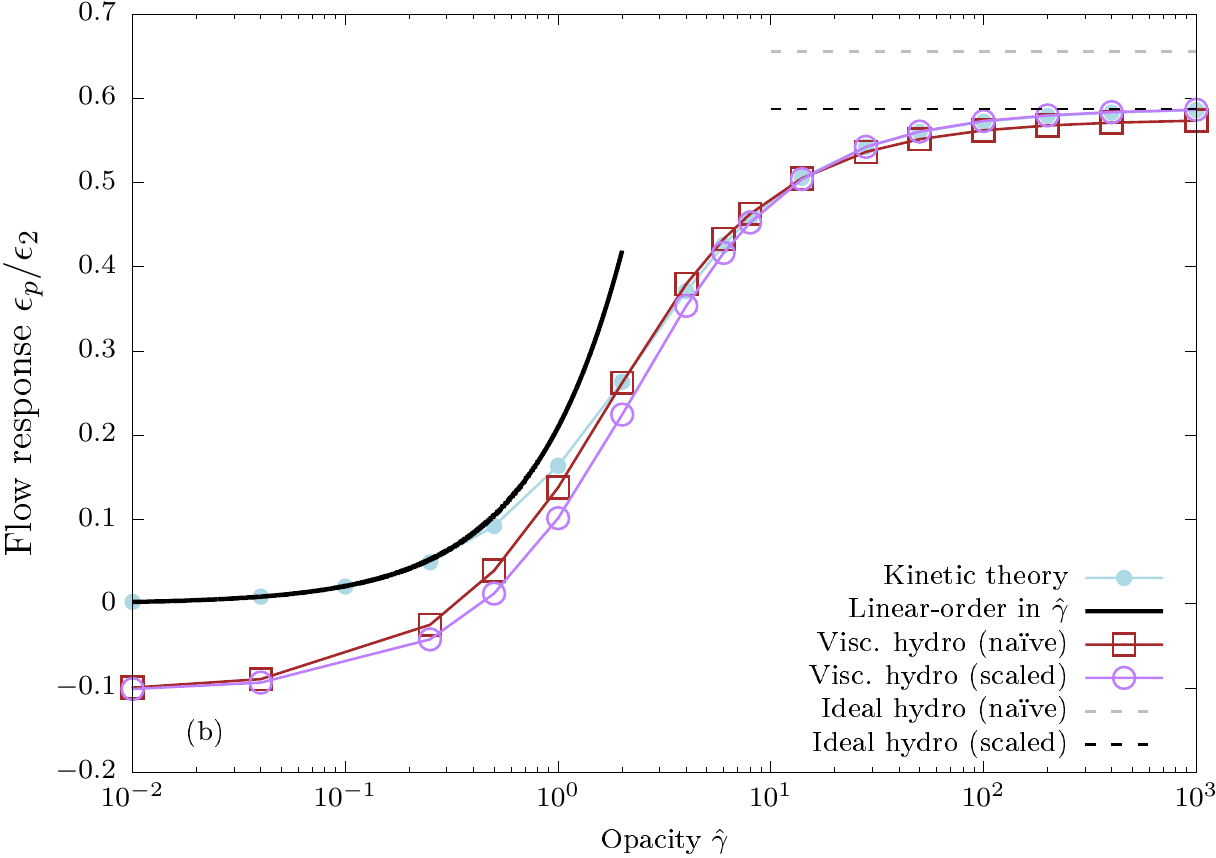}
\end{tabular}
}
\caption{Opacity dependence of (a) $dE_{\rm tr} / d\eta$ and (b) $\epsilon_p / \epsilon_2$ measured at $\tau = 4R$ obtained in kinetic theory (blue line with filled circles), viscous with na\"ive (brown line with empty squares) and scaled (purple line with empty circles). The ideal hydrodynamics limit for $\epsilon_p / \epsilon_2$ is shown with dotted gray and black lines for the na\"ive and scaled initializations, respectively.
}
\label{fig:scaled}
\end{figure}

As it turns out, there is some subtletly in defining the hydrodynamic analogue to a given setup in kinetic theory. Naively initializing hydrodynamics using the same energy profile used for the kinetic theory simulations leads to significant inconsistencies at the level of $dE_{\rm tr} / d\eta$, as shown in Fig.~\ref{fig:scaled}(a). Furthermore, the large $\hat{\gamma}$ limit of the elliptic flow coefficient $\epsilon_p/\epsilon_2$ differs in kinetic theory and hydrodynamics, and more importantly both differs from the prediction in ideal hydrodynamics. This is due to differences of the pre-equilibrium behaviour, as discussed in the following.

In the regime of large opacities $\hat{\gamma} \gg 1$ equilibration sets in before the transverse expansion and at early times the system can be effectively described as undergoing a homogeneous Bjorken flow at each point in the transverse plane. During the equilibration period, the system establishes a longitudinal pressure and begins to cool by performing work against the longitudinal expansion~\cite{Giacalone:2019ldn}. Because of the conformal behaviour, the hotter regions will undergo this equilibration faster than the colder ones, and hence the hotter regions also start to cool earlier than the colder regions.  Due to this {\it inhomogenous cooling}, the initial eccentricity decreases before the onset of the transverse expansion, and only a decreased spatial eccentricity will be converted to flow. Since the equilibration period is absent in ideal hydrodynamics, and described differently in viscous hydrodynamics and kinetic theory~\cite{Kurkela:2019set}, the effect on the eccentricity also differs between these descriptions. In Ref.~\cite{Ambrus:2021fej} we showed that the magnitudes of the eccentricity decrease match with the discrepancies arising in $\epsilon_p$.

%At early times we can regard the system as a collection of local Bjorken flows governed by their attractor solution~\cite{Giacalone:2019ldn,Kurkela:2019set}. For very large $\hat{\gamma}$, equilibration sets in during this period and we can use this approximaiton to study pre-equilibrium behaviour. In the early time limit, kinetic theory follows a free-streaming behaviour, keeping $\tau\epsilon$ constant, while in hydrodynamics the same quantity follows a positive power law. In both models, the evolution smoothly transitions to a $\tau^{-1/3}$ equilibrium power law. Because of the conformal behaviour, the hotter regions will undergo this process faster than the colder ones. Due to this inhomogeneous cooling, the initial eccentricity decreases before the onset of transverse expansion, and only the decreased eccentricity will be converted to flow. The total decrease is specific to the model and initial condition. In Ref.~\cite{Ambrus:2021fej} we showed that the magnitude of the eccentricity decrease matches with the discrepancies arising in $\epsilon_p$.

We will now present a modification of hydrodynamic setups that can alleviate the problem with pre-equilibrium. The idea we follow is to counteract the observed eccentricity decay by rescaling the hydro initial condition. As  $\tau\epsilon$ rises in magnitude in hydrodynamics, its initial condition is scaled down such that it will come into agreement with kinetic theory only after pre-equilibrium, assuming local Bjorken flow dynamics. As cooling proceeds inhomogeneously, the scaling factor is chosen locally, which will change the eccentricity in the hydro initial state such that it agrees with kinetic theory at later times. This is explained in more detail in~\cite{tobepublished}.

This scheme relies on the timescale separation between equilibration and transverse flow. At small opacities, equilibration will be interrupted by the onset of transverse expansion and cannot be encapsuled in the local Bjorken flow description. 
In these cases, we expect our scaling scheme to lead to unphysical results.
%In these cases, {\color{blue} \sout{our scaling scheme does not work well.} our proposed rescaling will lead to an overall decrease in the final-state energy, which becomes increasingly severe as $\hat{\gamma}$ is decreased.} 
% {\color{purple} One consequence of our proposed rescaling is an overall decrease in the final-state energy, which becomes increasingly severe as $\hat{\gamma}$ is decreased. Our scaling scheme does not work well.}

Fig.~\ref{fig:scaled} now compares the kinetic theory results for the opacity dependences of 
% transverse energy 
$dE_{\rm tr}/ d\eta$ and 
%elliptic flow 
$\epsilon_p$ with 
%and the scaled hydro scheme that we propose.
hydrodynamics results 
based on the naive and scaled initializations. %Despite the scaling, we identify the initial condition of kinetic theory to be the underlying truth and define the opacity $\hat{\gamma}$ and the normalization factors $\epn$ and $\epsilon_2$ on that basis.
For consistency, we characterize the simulation results based on the values of $\hat{\gamma}$, $dE^{(0)} / d\eta$ and $\epsilon_2$ corresponding to the kinetic theory setup.
% Due to the rise in energy during pre-equilibrium, naive hydro results severely overestimate $\frac{dE_{\rm tr}}{dy}$ at all opacities. Hydrodynamic results for $\epsilon_p$ at small opacities show an unphysical negative response. One would expect accurate results at large opacities, however, here the plot of naive hydro results shows the aforementioned discrepancies with kinetic theory.
%With the scaling scheme, both kinetic theory and viscous hydrodynamics seemlessly converge to the ideal hydrodynamics limit for $\epsilon_p$. 
When the initial conditions are scaled as discussed above, the viscous and ideal hydrodynamics results seemlessly converge to the kinetic theory results at large $\hat{\gamma}$.
%In the case of $\frac{dE_{\rm tr}}{d\eta}$, both schemes follow the same large opacity scaling law. 
%Going down in opacity, 
The kinetic theory and viscous hydrodynamics results stay in good agreement 
%up 
down to $\hat{\gamma}\gtrsim 10$. For small opacities, the hydrodynamics results for $\epsilon_p$ obtained using the scaled and naive initial conditions behave similarly.
%$dE_{\rm tr}/ d\eta$ is underestimated in scaled hydro, because the intial energy is scaled down according to the late time limit of cooling during the Bjorken scaling phase, which is never reached at small opacities, because this phase is interrupted by the onset of transverse expansion.
When the naive initialization is used, $dE_{\rm tr}/ d\eta$ increases as $\hat{\gamma}$ is decreased up to a plateau value, indicating that the time scale for transverse expansion becomes shorter than the pre-equilibrium one, such that the system never equilibrates. Since our scaling scheme reduces the initial energy density to account for the increase of $dE_{\rm tr} / d\eta$ over the whole free-streaming evolution, its final value will be increasingly underestimated as $\hat{\gamma}$ is decreased.

\section{Conclusions}

Our kinetic theory description is able to produce accurate results for cooling and flow for all opacitites, ranging from linear behaviour at small opacities to a saturation to ideal fluid behaviour at large opacities. In a naive comparison to hydrodynamics, we found sizeable discrepancies even at large opacities, which we determined to be due to different behaviour in the pre-equilibrium stage, where eccentricities decay by differing amounts. 

We presented a prescription to bring hydrodynamics into agreement with kinetic theory based on the idea of scaling the initial energy density in hydrodynamics in such a way that it agrees with kinetic theory only after the pre-equilibrium period in the case of purely longitudinal expansion. With this setup, we obtained agreement in the large opacity limit.

\textit{Acknowledgements:} This work is supported by the Deutsche Forschungsgemeinschaft (DFG, German Research Foundation)
through the CRC-TR 211 ``Strong-interaction matter under extreme conditions''– project
number 315477589 – TRR 211. V.E.A. was supported by a grant of the 
Ministry of Research, Innovation and Digitization, CNCS - UEFISCDI,
project number PN-III-P1-1.1-TE-2021-1707, within PNCDI III.
%Numerical calculations presented in this work were performed at Paderborn Center for Parallel Computing (PC2) and the Center for Scientific Computing (CSC) at the Goethe-University of Frankfurt.

%\vspace{-15pt}

%uncomment the following lines to place a figure
%\begin{figure}[htb]
%\centerline{%
%\includegraphics[width=12.5cm]{Fig1}}
%\caption{Plot of ...}
%\label{Fig:F2H}
%\end{figure}
\bibliographystyle{unsrtnt2}
\bibliography{references}

\end{document}